\documentclass[a4paper]{jpconf}
\usepackage[utf8]{inputenc}
\usepackage[english]{babel}
\usepackage{amsmath,amssymb}
\usepackage{textcomp,gensymb}
\usepackage{graphicx}
\usepackage{cite}

\newcommand{\be}{\begin{eqnarray}}
\newcommand{\ee}{\end{eqnarray}}

\begin{document}
\title{Monte Carlo studies of the Ising square lattice with competing interactions}

\author{A Kalz, A Honecker, S Fuchs and T Pruschke}

\address{Institut f\"ur Theoretische Physik,
        Georg-August-Universit\"at G\"ottingen, \\ 
	Friedrich-Hund-Platz 1, 37077 G\"ottingen, Germany}

\ead{kalz@theorie.physik.uni-goettingen.de}

\begin{abstract}
We use improved Monte-Carlo algorithms to study the antiferromagnetic 
2D-Ising model with competing interactions $J_1$ on nearest neighbour and $J_2$ on next-nearest neighbour bonds.
The finite-temperature phase diagram is divided by a critical point 
at $J_2 = J_1/2$  where the groundstate is highly degenerate. 
To analyse the phase boundaries we look at the specific heat
and the energy distribution for various ratios of $J_2/J_1$.
We find a first order transition for small $J_2 > J_1/2$ and the transition temperature
suppressed to $T_C=0$ at the critical point.
\end{abstract}

\section{Introduction \label{intro}}
The antiferromagnetic $J_1$-$J_2$ spin-1/2 Heisenberg model \cite{B:richter04,B:misguich05,P:mambrini06,P:richter08} 
and related bosonic models on the square lattice 
\cite{P:batrouni00,P:batrouni01,P:melko08,P:chen08}
with hopping $t_i$ and repulsion terms $J_i$ on nearest neighbour (NN) and next-nearest neighbour (NNN) bonds 
have been intensively studied over recent years via different approaches.
The treatment of the bosonic model via Quantum Monte-Carlo (QMC) algorithms suffers from freezing problems in the intermediate regime at $J_2 = J_1/2$ which can be traced to a groundstate degeneracy
of the static limit with hopping terms $t_i=0$. 
This motivated us to restudy the 2D-Ising model with competing interactions 
and solve these freezing problems via improved classical Monte-Carlo (MC) algorithms.
The 2D-Ising model has a long history and is still a topic of discussion \cite{P:krinsky79,P:landau80,P:lanbin80,P:lanbin85,P:bloete87,B:lanbin00};
especially the character of the phase boundary for $J_2 > J_1/2$ is an open question
\cite{P:lopez93,P:malakis06,P:monroe07,P:anjos08}. There are two scenarios discussed: Landau and Binder as well as
Malakis et al \cite{B:lanbin00,P:malakis06} used MC methods to infer a second order transition 
for all $J_2 > J_1/2$ with non universal exponents.
In contrast Lopez et al and Anjos et al use variational methods \cite{P:lopez93,P:anjos08} and 
find a first order transition for small $J_1 \gtrsim J_2 > J_1/2$ and a continuous transition only for larger $J_2$. 
Our simulation strengthens the latter scenario \cite{P:kalz08}.

We outline the model and the methods in section \ref{sec:model} and \ref{sec:methods}
and present our results in section \ref{sec:results}. We end with a short discussion and outlook.

\section{Model \label{sec:model}}
The model of hardcore bosons on a square lattice with competing interactions 
on NN bonds and NNN bonds is described by the hamiltonian:
\be
H_{\text{boson}} = - t_1\sum_{\text{NN}} (b^{\dagger}_i b_j + b_j^{\dagger} b_i) -
t_2\sum_{\text{NNN}} (b^{\dagger}_i b_j + b_j^{\dagger} b_i) 
+ V_1 \sum_{\text{NN}} n_i n_j + V_2\sum_{\text{NNN}} n_i n_j \label{e:fullham}
\ee
This hamiltonian can be mapped onto an eqivalent spin model which conincides 
with the $J_1$-$J_2$ spin-1/2 anisotropic Heisenberg model apart from the sign in front of the hopping terms $t_i$.
Taking the static limit $t_i = 0$ the hamiltonian reduces to that of the Ising model with competing antiferromagnetic
interactions $J_i > 0$:
\be
H_{\text{Ising}} = J_1 \sum_{\text{NN}} S_i S_j + J_2 \sum_{\text{NNN}} S_i S_j \,  ,
\quad S_i =\pm1 \label{e:isingham} \, .
\ee
We will study square lattices of linear extent $L$ with periodic
boundary conditions.

\begin{figure} % sketch of ordered states and degeneracy
\begin{minipage}{7.4cm}
\begin{center}
\resizebox{7.4cm}{!}{
  \includegraphics[]{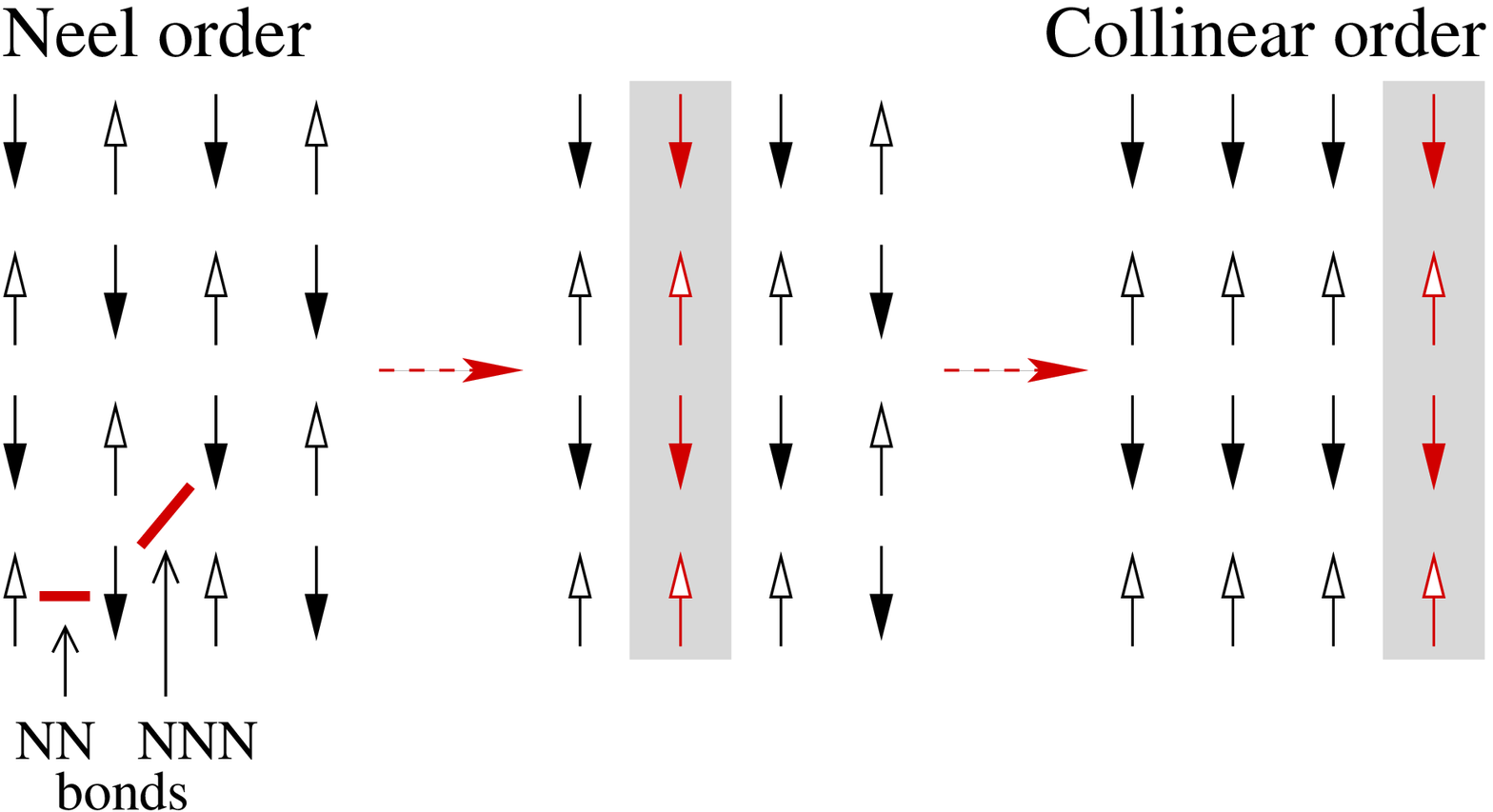}
}
\caption{Sketched are both ordered phases and
a third groundstate configuration at $J_2 = J_1/2$ (middle). Shaded areas mark flipped lines.}
\label{fig:degeneracy}
\end{center}
\end{minipage}
\hspace{1cm}
\begin{minipage}{7.4cm}
\begin{center}
\resizebox{7.4cm}{!}{%
  \includegraphics[]{binder_procee.eps}
}
\caption{Binder cumulants depending on temperature and $L$ for $J_2 = J_1$. In the inset the intersection area of the 
cumulants with errorbars gives $T_C/J_1 = 2.082 \pm 0.005$.}
\label{fig:cumulants} 
\end{center}
\end{minipage}
\end{figure}

For $J_2 < J_1/2$ the groundstate of (\ref{e:isingham}) is N\'eel-ordered (see Fig.~\ref{fig:degeneracy}, left) with energy:
\be
E_{\text{N\'eel}} = -2 N (J_1-J_2) \,,\quad N = L^2 \, .
\ee
For $J_2 > J_1/2$ the system orders in the collinear (or superantiferromagnetic)
phase (see Fig.~\ref{fig:degeneracy}, right). In this state every spin has two parallel and two antiparallel
aligned nearest neighbours. Thus, the energy depends only on the antiparallel diagonal bonds:
\be
E_{\text{Coll}} = -2 N J_2 \, .
\ee
At the critical point $J_2 = J_1/2$ separating these two phases
the transition temperature is suppressed and the
groundstate is highly degenerate of order $2^{L+1}-2$ (one of these states is shown 
in the middle panel of Fig.~\ref{fig:degeneracy}).
Due to this groundstate degeneracy the free energy is characterised by many local minima.
This leads to freezing problems for simple importance sampling MC simulations.

\section{Methods \label{sec:methods}}
In order to compute the transition temperature $T_C$ for ratios of $J_2/J_1$ 
very close to the critical point $J_2 = J_1/2$ we used parallel tempering
\cite{P:hukushima96,B:marinari98,P:hansmann97,P:katzgraber06} MC methods with additional 
line flip updates \cite{P:kalz08}.

To detect the finite temperature phase transitions we use the Binder cumulant (reduced fourth order cumulant)
$U_4 = 1 -\frac{\langle M^4\rangle}{3\langle M^2\rangle^2}$ 
\cite{P:binder81L,P:binder81Z,B:lanbin00} of the order parameter 
$M(\vec q) = \sqrt{S(\vec q)/N}$, where 
\be
S(\vec q)=\frac{1}{N}\sum_{i,j}e^{i\vec q\cdot(\vec x_i-\vec x_j)}\langle S_i  S_j \rangle \label{e:struct} 
\ee
is the respective structure factor (for an example see Fig.~\ref{fig:cumulants}). 

To determine the critical exponent $\nu$ we calculate the derivative of $U_4$ with respect to $T$
which is proportional to $L^{1/\nu}$ \cite{B:lanbin00}.

The behaviour of the specific heat and energy histograms \cite{P:challa86,P:berg91,P:borgs92,P:ngo08} 
for various ratios of $J_2/J_1$ provides further details to analyse the character 
of the given finite temperature phase boundaries and the transition at the critical point $J_2 = J_1/2$.

\section{Results \label{sec:results}}
\begin{figure} 
\begin{minipage}{7.4cm} % Phase diagram
\begin{center}
\resizebox{7.4cm}{!}{%
  \includegraphics[]{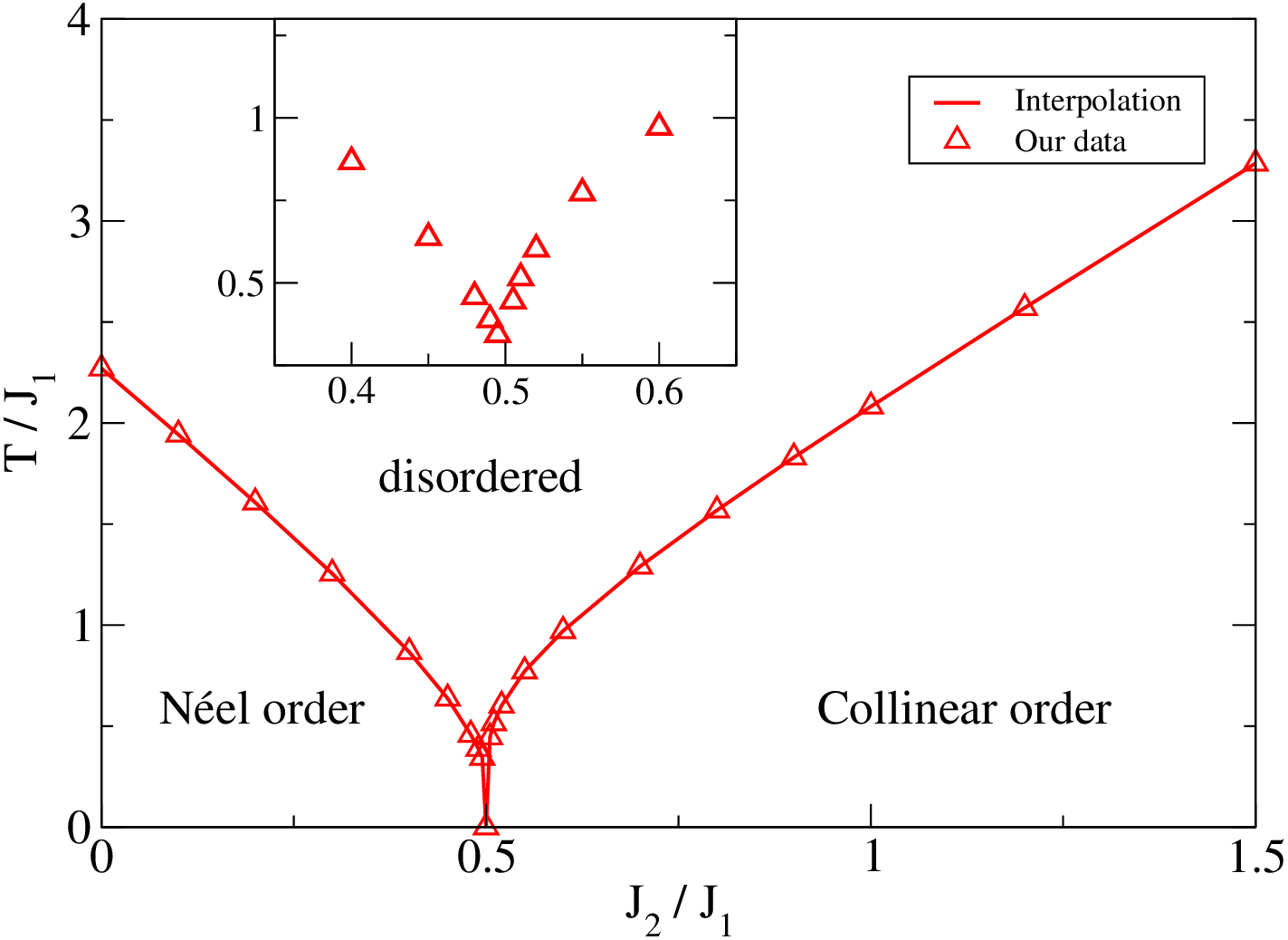}
}
\caption{Plotted is $T_C$ over the ratio $J_2/J_1$. Data were produced by \textit{parallel tempering 
MC simulations}.}
\label{fig:phase} 
\end{center}
\end{minipage}
\hspace{1cm}
\begin{minipage}{7.4cm} % histograms
\begin{center}
\resizebox{7.4cm}{!}{%
  \includegraphics[]{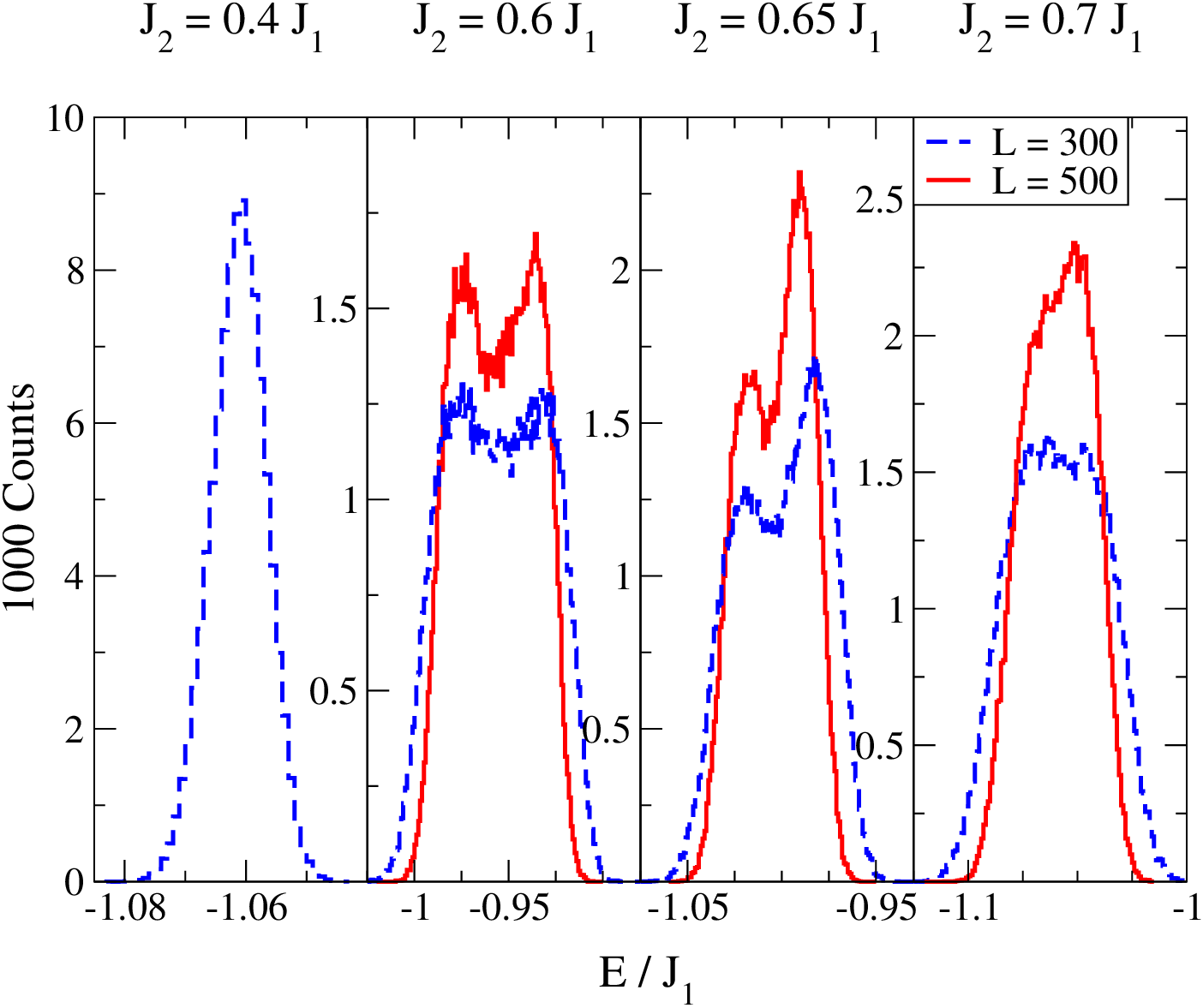}
}
\caption{Energy histogram for several ratios of $J_2/J_1$ at $T_C$ depending on given $L$.}
\label{fig:histo}
\end{center}
\end{minipage}
\end{figure}

In Fig.~\ref{fig:phase} we present the finite temperature phase diagram
for the classical Ising model (solid line) for various ratios $|J_2/J_1 - 0.5| \geq 0.005$ 
very close to the critical point.
The data is in good agreement with MC results from Landau and Binder for $|J_2/J_1 - 0.5| \geq 0.1$
\cite{P:lanbin85,P:landau80} and for $J_2 = J_1$ ($T_C/J_1 = 2.082 \pm 0.005$ as shown in 
Fig.~\ref{fig:cumulants}) with the recent result $T_C/J_1 = 2.0823 \pm 0.0017$ \cite{P:malakis06}.

Calculating lattice sizes up to $L=300$ we find the critical exponent $\nu = 1.002 \pm 0.010$ for $J_2 = 0.3J_1$. This and a slowly 
emerging peak in the specific heat confirms the assumption that for $J_2 < J_1/2$ the phase transition lies in the Ising universality 
class \cite{P:onsager44,P:fisher67}. For $J_2 > J_1/2$ the character of the phase transition is an open subject. We used a histogram method \cite{P:challa86,P:berg91,P:borgs92} to 
verify a first order transition \cite{P:ngo08} for small $J_2 > J_1/2$. The slowly emerging double peak in 
the energy distribution for $J_2 = 0.6~J_1$ and $J_2 = 0.65~J_1$ in Fig.~\ref{fig:histo} 
compared to the single gaussian peak for $J_2 < J_1/2$ indicates a first order transition. For larger $J_2$
the shape of the energy distribution only shows a broadened peak for the simulated lattice sizes. The calculation of $\nu$ for $J_2 > 
J_1/2$ strongly depends on the lattice size because of large crossover scales near the 
critical point and above. For $J_2 > J_1$ we find the critical exponent $\nu \approx 1$ 
($\nu = 1.02 \pm 0.02$ for $J_2 = 1.2~J_1$, $\nu = 0.96 \pm 0.02$ for $J_2 = 1.5~J_1$) for large lattice sizes 
which indicates that an Ising-like second order phase transition is recovered for $J_2 \gtrsim J_1$. 
So we believe that the character of the phase transition varies with $J_2$ and is transfered from a first order transition for 
$J_1/2 < J_2 \lesssim J_1$ to a continous transition for larger $J_2$ with universal exponents.

At the critical point $J_2 = J_1/2$ the finite size scaling analysis of the specific heat \cite{P:kalz08,P:berg91} 
suggests a suppression of the transition temperature to $T_C=0$.

\section{Discussion \label{sec:discussion}}
We used substantially enhanced MC methods to determine transition temperatures
and characterise the phase boundaries of the frustrated 2D-Ising model.
We solved the freezing problems which arise near the critical point at 
$J_2 = J_1/2$ due to a groundstate degeneracy and calculated critical temperatures 
in the direct vicinity. We found the phase transition for $J_2 < J_1/2$ to be continous 
and Ising-like. For $J_2 = J_1/2$ the phase transition is suppressed to zero temperature.
The character of the phase boundary on the right hand side of the phase diagram was discussed via histograms.
We decide the discussion about the character in favour of a first order transition for small 
$J_2 > J_1/2$ and find strong indications that the transition is transfered into a continous one 
with universal Ising exponents for large $J_2 > J_1$. However, this is complicated to show 
via MC simulations because of large crossover scales and the need for very large lattices.

Next we want to introduce finite hopping terms $t_i$ and study the bosonic model given in \eqref{e:fullham} 
using parallel tempering QMC simulations \cite{P:melko07}.

\ack{We acknowledge financial support by the Deutsche Forschungsgemeinschaft
under grant No.\ HO~2325/4-1 and through SFB602.}


\section*{References}
\begin{thebibliography}{9}
\bibitem{B:richter04} Richter J, Schulenburg J and Honecker A, in
\textit{Quantum Magnetism}, edited by Schollw\"ock U, Richter J, Farnell D J J and Bishop R F,
(Lecture Notes in Physics, 645) (Springer, Berlin, 2004) p.~85
\bibitem{B:misguich05} Misguich G and Lhuillier C, in \textit{Frustrated spin systems},  edited by Diep H T (World-Scientific, 2005)
\bibitem{P:mambrini06} Mambrini M, L\"auchli A, Poilblanc D and Mila F 2006,
\textit{Phys. Rev. B} \textbf{74} 144422
\bibitem{P:richter08} Darradi R, Derzhko O, Zinke R, Schulenburg J, Kr\"uger S E and Richter J 2008 Ground-state phases of the spin-1/2 
$J_1$-$J_2$ Heisenberg antiferromagnet on the square lattice: A high-order coupled cluster treatment \textit{Preprint arXiv:0806.3825}
\bibitem{P:batrouni00} Batrouni G G and Scalettar R T 2000 \textit{Phys. Rev. Lett.} \textbf{84} 1599
\bibitem{P:batrouni01} H\'ebert F, Batrouni G G, Scalettar R T,
Schmid G, Troyer M and Dorneich A 2001 \textit{Phys. Rev. B} \textbf{65} 014513
\bibitem{P:melko08} Chen Y C, Melko R G, Wessel S and Kao Y J 2008 \textit{Phys. Rev. B} \textbf{77} 014524
\bibitem{P:chen08} Ng K K and Chen Y C 2008 \textit{Phys. Rev. B} \textbf{77} 052506
\bibitem{P:krinsky79} Swendsen R H, Krinsky S 1979 \textit{Phys. Rev. Lett.} \textbf{43} 177
\bibitem{P:landau80} Landau D P 1980 \textit{Phys. Rev. B} \textbf{21} 1285
\bibitem{P:lanbin80} Binder K and Landau D P 1980 \textit{Phys. Rev. B} \textbf{21} 1941
\bibitem{P:lanbin85} Landau D P and Binder K 1985 \textit{Phys. Rev. B} \textbf{31} 5946
\bibitem{P:bloete87} Bl\"ote H W J, Compagner A and Hoogland A 1987 \textit{Physica} \textbf{141A} 375
\bibitem{B:lanbin00} Landau D P and Binder K, \textit{Monte Carlo Simulations in Statistical Physics} 
(Cambridge University Press, 2000)
\bibitem{P:lopez93} Mor\'an-L\'opez J L, Aguilera-Granja F and Sanchez J M 1993 \textit{Phys. Rev. B} \textbf{48} 3519
\bibitem{P:malakis06} Malakis A, Kalozoumis P and Tyraskis N 2006 \textit{Eur. Phys. J. B} \textbf{50} 63
\bibitem{P:monroe07} Monroe J L and Kim S 2007 \textit{Phys. Rev. E} \textbf{76} 021123
\bibitem{P:anjos08} dos Anjos R A, Viana J R and de Sousa J R 2008 \textit{Phys. Lett. A} \textbf{372} 1180
\bibitem{P:kalz08} Kalz A, Honecker A, Fuchs S and Pruschke T 2008 Phase diagram of the Ising square lattice with competing interactions \textit{Preprint arXiv:0805.0983} (\emph{to appear in Eur. Phys. J. B})
\bibitem{P:hukushima96} Hukushima K and Nemoto K 1996 \textit{J. Phys. Soc. Jpn.} \textbf{65} 1604
\bibitem{B:marinari98} Marinari E, Lecture Notes in Physics, vol. 501 (Springer, 1998) [\textbf{cond-mat/9612010}]
\bibitem{P:hansmann97} Hansmann U H E 1997 \textit{Chem. Phys. Lett.} \textbf{281} 140
\bibitem{P:katzgraber06} Katzgraber H G, Trebst S, Huse D A and Troyer M 2006 \textit{J. Stat. Mech.} P03018
\bibitem{P:binder81L} Binder K 1981 \textit{Phys. Rev. Lett.} \textbf{47} 693
\bibitem{P:binder81Z} Binder K 1981 \textit{Z. Phys. B} \textbf{43} 119
\bibitem{P:challa86} Challa M S S, Landau D P and Binder K 1986 \textit{Phys. Rev. B} \textbf{34} 1841
\bibitem{P:berg91} Alves N A, Berg B A and Villanova R 1991 \textit{Phys. Rev. B} \textbf{43} 5846
\bibitem{P:borgs92} Borgs C and Janke W 1992 \textit{Phys. Rev. Lett.} \textbf{68} 1738
\bibitem{P:ngo08} Ngo V T and Diep H T 2008 Phase Transition in Heisenberg Stacked Triangular Antiferromagnets: End of a Controversy \textit{Preprint arXiv:0808.0520}
\bibitem{P:onsager44} Onsager L 1944 \textit{Phys. Rev.} \textbf{65} 117
\bibitem{P:fisher67} Fisher M E and Burford R J 1967 \textit{Phys. Rev.} \textbf{156} 583
\bibitem{P:melko07} Melko R G 2007 \textit{J. Phys.: Condens. Matter} \textbf{19} 145203
\end{thebibliography}
\end{document}